\documentclass[%
 reprint,
 amsmath,amssymb,
aip,
]{revtex4-1}
\usepackage{natbib}
\usepackage{graphicx}
\usepackage{dcolumn}
\usepackage{bm}
\usepackage{braket}
\usepackage{url}
\usepackage{caption}
\usepackage{subcaption}
\usepackage[colorlinks,citecolor={blue}]{hyperref}

\usepackage[colorinlistoftodos]{todonotes}
\begin{document}

\preprint{AIP/}

\title{A study of the dense Uniform Electron Gas with high orders of Coupled Cluster}

\author{Verena A. Neufeld}
 \email{van26@cam.ac.uk}
\author{Alex J. W. Thom}%
\affiliation{%
 Department of Chemistry, Lensfield Road, Cambridge, CB2 1EW, United Kingdom
}%

\date{\today}

\begin{abstract}
We investigate the accuracies of different coupled cluster levels in a
finite model solid, the 14 electron spin-non-polarised uniform electron gas.
For densities between
$r_{\mathrm{s}} =$ 0.5 $\mathrm{a}_\mathrm{0}$ and $r_{\mathrm{s}} =$ 5
$\mathrm{a}_\mathrm{0}$, we
calculate ground state correlation energies with stochastic coupled cluster
ranging from coupled cluster singles and doubles (CCSD) to coupled cluster
including all excitations up to quintuples (CCSDTQ5). We find the need to add triple excitations for an
accuracy of 0.01eV/electron beyond $r_{\mathrm{s}} =$ 0.5
$\mathrm{a}_\mathrm{0}$. Quadruple excitations start being significant past
$r_{\mathrm{s}} =$ 3 $\mathrm{a}_\mathrm{0}$. At $r_{\mathrm{s}} =$ 5
$\mathrm{a}_\mathrm{0}$, CCSD gives a correlation energy with a 16\% error and
CCSDT is in error by 2\% compared to the CCSDTQ5 result. CCSDTQ5 gives an energy
in agreement with full configuration interaction quantum Monte Carlo results.
\end{abstract}

\maketitle

\section{Introduction}
Coupled cluster theory \cite{Coester1960,Cizek1966,Cizek1971} is known as the
gold standard of \textit{ab initio} molecular simulations giving energies to
chemical accuracy of about 1 kcal mol$^{-1}$ (see review Ref.
\onlinecite{Bartlett2007}). Moreover, its accuracy is systematically improvable
as more excitation levels are added. Driven by the need for systematically
improvable methods in solids, coupled cluster is now increasingly being applied
to periodic systems, see e.g. Refs.
\onlinecite{Hirata2001a,Hirata2004,Gruneis2011,Booth2013,Gruneis2015,
Gruneis2015a,Liao2016,McClain2017c}.
In this paper we apply the stochastic coupled cluster method
\cite{Thom2010,Spencer2016} to the dense uniform electron gas to assess the
performance of coupled cluster in periodic systems at representative electron
densities. Performing coupled cluster stochastically can often reduce the memory
requirements and computational scaling. It can therefore reach higher basis sets
and coupled cluster levels than conventional implementations.
\par While coupled cluster is just starting to emerge as a useful tool for solid
calculations, density functional theory (DFT) \cite{Kohn1965,Hohenberg1964} is
one of the most widely used \textit{ab initio} electronic structure methods in
extended systems. It scales favourably ($\mathcal{O}$($n^3$) or even
$\mathcal{O}$($n$) (see e.g. Ref. \onlinecite{Skylaris2005}) where $n$ is a
measure for the system size) but it is not systematically improvable and it can
have difficulties with strongly correlated systems \cite{Cohen2012}. These and
other shortcomings have led to an interest in applying alternative methods to
periodic systems. Besides coupled cluster, these include for example the random
phase approximation (RPA) \cite{Bohm1951,Pines1952,Bohm1953}, second order
M{\o}ller-Plesset perturbation theory (MP2) \cite{Moller1934}, diffusion
Monte Carlo (DMC) \cite{Foulkes2001} and (initiator) full configuration
interaction quantum Monte Carlo (FCIQMC) \cite{Booth2009,Cleland2010,Booth2013}.
RPA yields a ground state energy that is equal to the energy output of a version
of ring diagram coupled cluster doubles (CCD) \cite{Freeman1977,Scuseria2008}.
MP2 has been shown to diverge in the thermodynamic limit in the uniform electron
gas whereas it is unclear whether coupled cluster singles and doubles (CCSD)
does \cite{Shepherd2012d,Shepherd2013}. DMC scales as $\mathcal{O}$($n^3$)
\cite{Foulkes2001} and can give the exact answer provided sufficient \textit{a
priori} information about the wavefunction is available. FCIQMC gives exact
energies within a finite basis set without requiring \textit{a priori} knowledge
of the wavefunction. Most such calculations are now performed with the initiator
approximation \cite{Cleland2010} that adds a bias to the energy which can be
systematically reduced as more Monte Carlo particles are added to the system.
FCIQMC has so far been applied to small systems, including LiH
(3$\times$3$\times$3 $k$ point mesh with 2 electrons and 2 orbitals per $k$
point) and diamond (2$\times$2$\times$2 $k$ points with 8 electrons and 8
orbitals per $k$ point) \cite{Booth2013}. Since FCIQMC samples the whole Hilbert
space, it is often more expensive than a level of coupled cluster that is
sufficient for accurate energies.
\par The system we study here is the uniform electron gas (UEG)
\cite{MartinUEGChapter,Giuliani2005,Loos2016a} which is a simple model for a
periodic system where the positive lattice potential of the atomic nuclei is
approximated by a uniform positive background potential; the energies of electron gases play
an important role in density functional theory
\cite{Giuliani2005,Giuliani2005dft,Loos2016a}. There exists accurate
ground state energy data for the high density regime based on the finite UEG
with the FCIQMC \cite{Shepherd2012,Shepherd2012a,Shepherd2012b} and DMC
\cite{Ceperley1980,Ortiz1994,Ortiz1997,Ortiz1999,Kwon1998,Holzmann2003,
LopezRios2006,Drummond2008,Spink2013} methods. Versions of coupled cluster
have been applied to the UEG in the thermodynamic limit, see e.g.
\cite{Freeman1977,Bishop1978,Bishop1982}.
CCSD and CCSDT have been applied to the finite three-dimensional (3D) UEG
\cite{Shepherd2012,Shepherd2013,Roggero2013,Spencer2016,McClain2016a,Shepherd2016a}.
Shepherd \cite{Shepherd2016a} has extrapolated finite CCSD/CCD results in the 3D UEG
to the thermodynamic limit and has compared them to Ceperley and Alder's DMC
energies \cite{Ceperley1980} (see figure 2c in Ref. \onlinecite{Shepherd2016a}). Using
these DMC energies as a reference, the extrapolated CCSD correlation energy has
an error of under 10\% at $r_{\mathrm{s}} =$ 1.0  $\mathrm{a}_\mathrm{0}$ which
increases to about 20\% at $r_{\mathrm{s}} =$ 5.0 $\mathrm{a}_\mathrm{0}$.
Another recent study \cite{Spencer2016} has performed
initiator and non-initiator stochastic coupled cluster in the CCSD and CCSDT
levels on the dense 14 electron 3D UEG. The difference between CCSD and CCSDT
was found to be significant even in the low correlation regime at
$r_{\mathrm{s}} <$ 1.0 $\mathrm{a}_\mathrm{0}$. $r_{\mathrm{s}}$ is the radius
of a sphere that on average contains one electron. In this paper, we apply
coupled cluster up to the CCSDTQ5 level which included quintuple excitations
directly to the 14 electron non-spin-polarized UEG in the range $r_{\mathrm{s}} = $ 0.5 to 5.0
$\mathrm{a}_\mathrm{0}$ which is representative of some common simple solids
(e.g. see Ref. \onlinecite{MartinUEGChapter}). We compare with (initiator)
FCIQMC \cite{Shepherd2012b} and MP2 \cite{Shepherd2012} results. Using coupled
cluster levels from CCSD to CCSDTQ5, we aim to answer the question what coupled
cluster level is needed to accurately model simple finite solids with certain
densities, represented by the $r_{\mathrm{s}}$ parameter, with coupled cluster.

\section{Coupled Cluster Monte Carlo}
Coupled Cluster Monte Carlo (CCMC) \cite{Thom2010,Spencer2016} is a stochastic
version of the coupled cluster method
\cite{Coester1960,Cizek1966,Cizek1971,Bartlett2007}. The energies obtained are
consistent with conventional coupled cluster while often saving computational and
memory cost \cite{Thom2010,Spencer2016,Scott2017b}. Recent developments include
linked CCMC \cite{Franklin2016}, the initiator approximation for CCMC
\cite{Spencer2016} and the even selection feature \cite{Scott2017b}. This section
gives a brief overview over the method which is described more thoroughly in the
literature \cite{Thom2010,Spencer2016}.
\par Coupled cluster theory solves the Schr{\"o}dinger equation for the ground
state energy. The ground state wavefunction $\Psi_0$ is constructed from the
reference wavefunction $\Psi_{\mathrm{ref}}$ using the ansatz
\begin{equation}
\Psi_0 \propto \exp(\hat{T}) \Psi_{\mathrm{ref}},
\label{eq:ccmc_groundwavefunc}
\end{equation}
where $\hat{T} =
\sum_{\mathbf{i}}t_{\mathbf{i}}\hat{a}_{\mathbf{i}}$.
Wavefunctions are expressed in a Slater determinant $D_{\mathbf{i}}$
basis. In this study, $\Psi_{\mathrm{ref}} = D_{\mathbf{0}}$, the
Hartree-Fock Slater determinant. $\hat{a}_{\mathbf{i}}$ are excitors, that
produce excited Slater determinants $\hat{a}_{\mathbf{i}}
D_{\mathbf{0}} = D_{\mathbf{i}}$. $t_{\mathbf{i}}$ are the
corresponding coefficients of $\hat{a}_{\mathbf{i}}$. If the sum is over
all possible $\hat{a}_{\mathbf{i}}$, $\Psi_0$ will tend to the full
configuration interaction (FCI) wavefunction. Coupled cluster theory has the
advantage over doing (F)CI that it can truncate the sum for $\hat{T}$ to only
include some excitors $\hat{a}_{\mathbf{i}}$ while still being size
consistent. Coupled cluster singles and doubles (CCSD) for example only includes
excitors that excite one or two electrons whereas CCSDT also includes excitors
that excite three electrons from the reference and so on. Due to the exponential
in equation \ref{eq:ccmc_groundwavefunc}, higher order excitations are still
present indirectly, created by a combination of lower order ones and therefore
dependent on their coefficients $t_{\mathbf{i}}$. \par Stochastic Coupled
Cluster makes use of the sparsity of the wavefunction, and uses sampling to
decrease computational and memory costs. To understand the sampling algorithm,
we first project the Schr{\"o}dinger equation onto some determinant
$\bra{D_{\mathbf{m}}}$ giving a set of equations
\begin{equation}
	\bra{D_{\mathbf{m}}} \hat{H} - E \ket{\Psi_0} = 0.
	\label{eq:ccmc_Schroedinger_1}
\end{equation}
Instead of explicitly solving these equations, the ground state wavefunction is
formed by a projection from the reference, $\Psi_0=\exp{\left(-\tau \hat
H\right)}\Psi_{\mathrm{ref}}$, where imaginary time, ${\tau\rightarrow\infty}$.
After some manipulation, \cite{Thom2010,Spencer2016} this yields an iterative
equation for the amplitudes $t_{\mathbf{i}}$,
\begin{equation}
t_{\mathbf{i}}(\tau+\delta\tau) = t_{\mathbf{i}}(\tau) - \delta \tau \bra{D_{\mathbf{i}}} \hat{H} - E \ket{\Psi(\tau)}.
\label{eq:ccmc_iterative_t}
\end{equation}
Monte Carlo particles are placed on the excitors $\hat{a}_{\mathbf{i}}$.
They are then propagated to sample equation \ref{eq:ccmc_iterative_t} which is
explained in more detail in the following paragraph. At convergence, the average
population on an excitor $\hat{a}_{\mathbf{i}}$ corresponds to its
coefficient $t_{\mathbf{i}}$. These particles do not have to be discrete
and can take real-valued weights \cite{Petruzielo2012,Overy2014}.
\par We start the Monte Carlo sampling by randomly picking a cluster (i.e. a
combination) of excitors that are occupied by particles. They act on the
reference determinant to yield an excited determinant $D_{\mathbf{n}}$.
The three major steps are \cite{Thom2010,Booth2009}:
\begin{itemize}
	\item \textit{Spawn:} Another determinant $D_{\mathbf{m}}$ that may be
	unoccupied or occupied is randomly chosen. With a probability proportional to
	$|\bra{D_{\mathbf{m}}} \hat{H} \ket{D_{\mathbf{n}}}|$, Monte
	Carlo particles can spawn to $\hat{a}_{\mathbf{m}}$.
	\item \textit{Death/Birth:} With a probability proportional to
	$|\bra{D_{\mathbf{n}}} \hat{H} - S - E_{\mathrm{HF}}
	\ket{D_{\mathbf{n}}}|$ a particle is placed on $a_{\mathbf{n}}$. $S$ is the
	population-controlling shift, described below, and $E_{\mathrm{HF}}$ is the
	Hartree-Fock energy.
	\item \textit{Annihilation:} Finally, particles of opposite sign on the same
	excitor are removed.
\end{itemize}
The ground state correlation energy is estimated by the projected energy
\begin{equation}
E_{\mathrm{proj.}}(\tau) = \frac{\bra{D_{\mathbf{0}}} \hat{H} - E_{\mathrm{HF}}
\ket{\Psi(\tau)}}{\braket{D_{\mathbf{0}}|\Psi(\tau)}}
\label{eq:projE}
\end{equation}
and, independently, by the above-mentioned shift. The shift is usually set to
zero at the beginning and when the particle number $N$ is high enough (when we
have passed the plateau phase of the sampling \cite{Thom2010,Spencer2012}) it is
varied as \cite{Booth2009}
\begin{equation}
S(\tau)=S(\tau - Z\delta\tau) - \frac{\gamma}{Z\delta\tau}
\ln\left(\frac{N(\tau)}{N(\tau-Z\delta\tau)}\right)
\end{equation}
where $\gamma$ is a damping parameter and $Z$ is the number of iterations to
pass before the shift is updated.
\par Franklin et al. \cite{Franklin2016} have modified equation \ref{eq:ccmc_iterative_t} (which later had $E$ replaced by the sum of the shift $S$ and the Hartree Fock energy $E_{\mathrm{HF}}$) to
\begin{equation}
\begin{split}
t_{\mathbf{i}}(\tau+\delta\tau) = t_{\mathbf{i}}(\tau)  - \delta \tau
\bra{D_{\mathbf{i}}} \hat{H} - E_{\mathrm{proj.}} - E_{\mathrm{HF}} \ket{\Psi(\tau)} -\\
\delta \tau(E_{\mathrm{proj.}}-S)t_{\mathbf{i}}(\tau),
\label{eq:ccmc_iterative_t_2}
\end{split}
\end{equation}
which we use as well. We use equation \ref{eq:projE} to find an estimate of for
$E_{\mathrm{proj.}}$. This change does not affect
the \textit{Spawn} and \textit{Annihilation} steps. If a single excitor
$a_{\mathbf{n}}$ was selected before the \textit{Spawn} step, the (modified)
\textit{Death/Birth} step causes a particle to die/be created on $a_{\mathbf{n}}$
with a probability proportional to $|\bra{D_{\mathbf{n}}} \hat{H} 
- S - E_{\mathrm{HF}} \ket{D_{\mathbf{n}}}|$. For composite
clusters, i.e. if two or more excitors were selected and collapsed to
$a_{\mathbf{n}}$, the probability is proportional to
$|\bra{D_{\mathbf{n}}} \hat{H} - E_{\mathrm{proj.}}(\tau) - E_{\mathrm{HF}} 
\ket{D_{\mathbf{n}}}|$ instead as we do not sample the third term on the right
hand side of equation \ref{eq:ccmc_iterative_t_2} then.
\par For our stochastic calculations, we have made use of development versions
of the HANDE code \footnote{See Ref. \onlinecite{HANDEpaper} and
\url{http://www.hande.org.uk/} for information and code.}. We have used the
cluster multispawn feature \cite{parallelPaper} and the full non-composite
cluster selection described in Ref. \onlinecite{parallelPaper} using one MPI process
divided up into OpenMP threads when running CCMC. We have also run some FCIQMC
calculations to compare our CCMC results to and we used the conventional and
initiator versions
for FCIQMC \cite{Booth2009,Cleland2010} while only using non-initiator CCMC. The
error bars of the data presented here were estimated by reblocking analysis
\cite{Flyvbjerg1989} using pyblock \footnote{See
\url{https://github.com/jsspencer/pyblock} for information and code.} and the
correlation energies are obtained from the projected energy. For the data
presented here, the projected energy agrees with the shift within 2$\sigma$.
Errors were combined in quadrature. We found no significant population control
bias using a reweighting scheme used in DMC \cite{Umrigar1993} and adapted to
FCIQMC \cite{Vigor2015}.

\section{Uniform Electron Gas}
We used a plane wave basis and studied the 14 electron non-spin-polarised electron gas. The simulation was performed in three dimensional
$\bm{k}$ space, where the set of $\bm{k}$ are the wavevectors of the $M/2$ plane
waves, with a cubic simulation box with sides of length $L$. A kinetic energy
cutoff was used to select the plane waves. In $\bm{k}$ space and using second
quantisation, the Hamiltonian is expressed as
\begin{equation}
\begin{split}
\hat{H} = \sum_{\bm{k}\vphantom{\bm{'}}} \frac{1}{2} (\bm{k}^2 +
V_{\mathrm{M^{\vphantom{\dagger}}}\vphantom{'}})
\hat{c}^{\dagger}_{\bm{k}\vphantom{\bm{'}}}
\hat{c}_{\bm{k}\vphantom{\bm{'}}}^{\vphantom{\dagger}} + 
\sum_{\bm{q}\neq
\bm{0},\bm{k},\bm{k'}} \frac{1}{2} \frac{4\pi}{|\bm{q}|^2L^3}
\hat{c}^{\dagger}_{\bm{k}\vphantom{\bm{'}} + \bm{q}} \hat{c}^{\dagger}_{\bm{k'} +
\bm{q}} \hat{c}^{\vphantom{\dagger}}_{\bm{k}\vphantom{\bm{'}}}
\hat{c}^{\vphantom{\dagger}}_{\bm{k'}}.
\end{split}
\end{equation}
$\hat{c}^{\dagger}_{\bm{k}}$/$\hat{c}^{\vphantom{\dagger}}_{\bm{k}}$
creates/annihilates an electron with momentum $\bm{k}$ and
$V_{\mathrm{M^{\vphantom{\dagger}}}\vphantom{'}}$ is the Madelung constant that
does not affect the correlation energy.
$r_{\mathrm{s}} = (\frac{3L^3}{4\pi N})^{\frac{1}{3}}$ where $N$ is the
number of electrons.

\section{Extrapolation to Complete Basis Set Limit}
Coupled cluster singles and doubles (CCSD) is the least expensive level of
coupled cluster. Owing to momentum and spin conservation, CCSD is equivalent to
CCD in the UEG. At first, we extrapolated CCSD calculations to the complete
basis set (CBS) limit for the 14 electron UEG. We then estimated the CBS limit
of the other truncation levels studied by extrapolating energy differences
between truncation levels and adding this to the CBS CCSD result. This is
similar to the idea of focal point analysis as described in e.g. Ref.
\onlinecite{East1993}.
\par Shepherd et al. \cite{Shepherd2012} have shown that for MP2, the correlation
energy for a finite basis set with $M$ spinorbitals goes as $1/M$ in the
leading order for large $M$. They and other studies
\cite{Shepherd2012a,Shepherd2012b,Shepherd2014,Shepherd2014a,Shepherd2016a,
Spencer2016} have used this trend and shown that it also holds reasonably well
for CCSD and FCI(QMC). These studies have usually excluded points with larger
$1/M$ that were no longer in the region in which $1/M$ is a good fit.
\par In this study, we have decided to modify this approach to allow higher
orders of $1/M$ to be considered as well. This accounts for the fact that $1/M$
is merely a leading order term and by adding higher orders we allow for
correction terms to account for the part of the energy not accounted for by
$1/M$. There are two aspects that need to be considered
when choosing the best fit curve: What polynomial are we fitting, i.e. what is
the highest order of $1/M$ to include, and how many points with high $1/M$
should be excluded from the fit?
\par Starting with the lowest order polynomial to
fit ($1/M$ when fitting CCSD and a constant when fitting coupled cluster differences),
we first fit all the
data points and then start excluding points with lowest $M$. For each fit, we
calculate $\chi^2$ over number of degrees of freedom \#d.o.f..
$\chi^2 = \sum_i \left(\frac{f(x_i) - y_i}{\sigma_i}\right)^2$
where $y_i$ is a data value, $f(x_i)$ is its fitted value and $\sigma_i$ is the
standard deviation of $y_i$ \cite{McPherson1990}. As soon as we reach a local
minimum in the $\chi^2$/\#d.o.f. value, we stop removing
points and note down the value at $1/M$ = 0 given by the fit at the local minimum.
If no local minimum can be found before there are as few data points left as the
number of fitting parameters, then the search for a best fit for the first
polynomial was unsuccessful. We then repeat this procedure of consecutively
removing data points with the next order polynomials, initially starting with a
full set of data points again. We fit linear, quadratic and cubic polynomials
and a constant as well if we are fitting to differences. Finally, we compare
the results of the fits at local minima in the number of points at $1/M$ = 0.
If the lowest order fit result agrees with the higher order ones within 2$\sigma$,
we accept it as the CBS result. If it does not agree with all the higher ones,
we compare the second lowest order fit result to
its higher order fit results, etc. This process can continue up comparing the
CBS results from the highest two polynomials. If there is still no CBS result at the end, then the
extrapolation was not successful and a CBS value has to be estimated (see
results section for individual cases).
\begin{figure}
\centering
	\includegraphics[width=8.5cm,keepaspectratio]{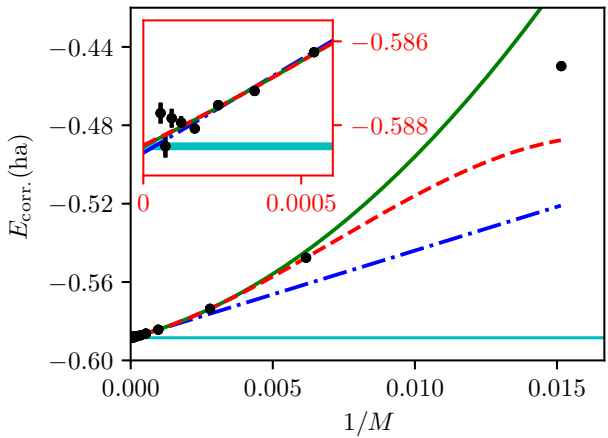}
	\caption{Extrapolating correlation energy against $1/M$ for $r_{\mathrm{s}}$ =
	0.5 $\mathrm{a}_\mathrm{0}$ CCSD and 14 electrons with the best fit linear line
	(blue, dashed, excluding three data points) giving $b_0$ = -0.58866(5) ha, best
	quadratic fit (green, solid line, excluding two data points) with $b_0$ =
	-0.58850(6) ha and best cubic fit (red, long dashes, excluding one data point),
	giving $b_0$ = -0.58848(7) ha. The CBS limit is then taken to be -0.58850(6) ha
	from the quadratic fit, as the linear fit and the quadratic fit do not agree
	within 2$\sigma$ whereas the quadratic and cubic fits agree within 2$\sigma$.
	The CBS result is shown with a light blue horizontal line that has a thickness
	of twice its error.}
	\label{fig:extrapolation}
\end{figure}
\par As an example, figure \ref{fig:extrapolation} shows the best fits with the
lowest $\chi^2/$\#d.o.f. for $r_{\mathrm{s}}$ = 0.5 $\mathrm{a}_\mathrm{0}$
CCSD and 14 electrons. The linear and the quadratic fit intercepts do not agree
within 2$\sigma$. The quadratic and cubic fits agree which meant that we took
the quadratic fit intercept as the CBS result. We have used the curve\_fit
function in the SciPy \footnote{E. Jones, T. Oliphant, P. Peterson et al.,
\textit{SciPy: Open source scientific tools for Python}. See
\url{https://www.scipy.org/} for more information.} optimize module for curve
fitting and Matplotlib \cite{Hunter2007} for plotting. The standard errors of
the correlation energy were treated as absolute and not relative weights.

\section{Results}
\begin{figure}
\centering
	\begin{subfigure}[h]{8.5cm}
	\includegraphics[width=1.0\linewidth,keepaspectratio]{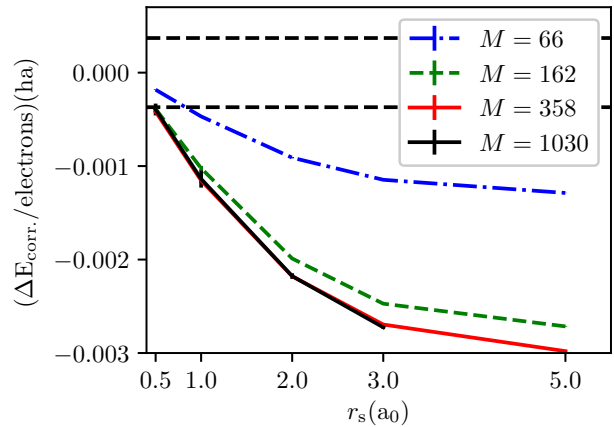}
	\caption{CCSD - CCSDT correlation energy difference}
	\label{fig:energydiffs1}
	\end{subfigure}
	\newline
	\begin{subfigure}[h]{8.5cm}
	\includegraphics[width=1.0\linewidth,keepaspectratio]{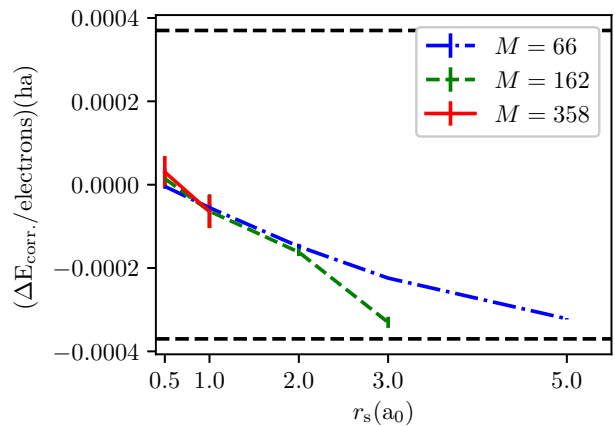}
	\caption{CCSDT - CCSDTQ correlation energy difference}
	\label{fig:energydiffs2}
	\end{subfigure}
	\newline
	\begin{subfigure}[h]{8.5cm}
	\includegraphics[width=1.0\linewidth,keepaspectratio]{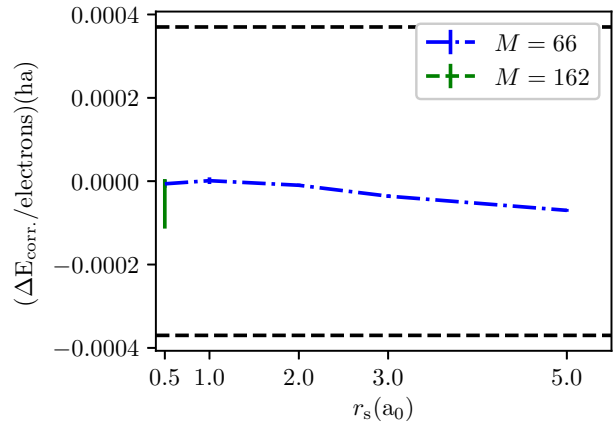}
	\caption{CCSDTQ - CCSDTQ5 correlation energy difference}
	\label{fig:energydiffs3}
	\end{subfigure}
	\newline
	\caption{Coupled cluster energy per electron differences at spinorbitals $M$ 
	= 66, 162, 358, 1030. The dashed horizontal lines show an accuracy of 0.01
	eV/electron.}
	\label{fig:energydiffs}
\end{figure}

Figure \ref{fig:energydiffs} shows how the differences in correlation energy
between consecutive coupled cluster levels vary with $r_{\mathrm{s}}$ for
different numbers of spinorbitals $M$. As a reference, an accuracy of 0.01
eV/electron = 0.00037 ha/electron is shown with dashed horizontal lines.
This is of a similar order of magnitude as chemical accuracy (ca. 0.04
eV/molecule \cite{Foulkes2001}). 
To distinguish
solid phases from each other, enthalpy differences of about 0.1 eV/atom often
need to be resolved and at room temperature an accuracy of 0.01 eV in the
energy is desired (see Ref. \onlinecite{Wagner2016} for details). We
have therefore chosen 0.01 eV/electron as a guide for energies to be of sufficient accuracy.
\par The CCSD to CCSDTQ5 CBS values are summarized in table \ref{tab:summary}.
Note that while figure \ref{fig:energydiffs} quotes energies in energies per
electron, table \ref{tab:summary} shows energies for 14 electrons. First, the
CCSD CBS value was found and then the CBS limit of differences between
consecutive coupled cluster levels were added on to find the CBS limit of the
other truncation levels. For $r_{\mathrm{s}}$ up to 2.0 $\mathrm{a}_\mathrm{0}$,
earlier CCSD and CCSDT results \cite{Spencer2016} are shown as well. MP2 results
\cite{Shepherd2012} and FCIQMC are given for comparison. For $r_{\mathrm{s}}$ =
0.5, 1.0, 2.0 and 5.0 $\mathrm{a}_\mathrm{0}$ FCIQMC values from Shepherd et al.
\cite{Shepherd2012b} are given and additionally for $r_{\mathrm{s}}$ = 0.5 and
1.0 $\mathrm{a}_\mathrm{0}$, new FCIQMC CBS results are presented for
comparison. When using the initiator approximation \cite{Cleland2010}, the
FCIQMC correlation energies values for a certain number of spinorbitals $M$ were
estimated by fitting horizontal lines to energy against number of Monte Carlo
particles curves, consecutively removing data points with the least number of
particles. The energy at the global minimum in $\chi^2/$\#d.o.f. when fitting a
horizontal line is taken as the energy result. The error in the average number
of particles was very small and therefore ignored. For the (\textit{i})FCIQMC
results with $r_{\mathrm{s}}$ = 0.5 and 1.0 $\mathrm{a}_\mathrm{0}$, the
initiator approximation was used for $M$ greater then 358 and 66 respectively.
The initiator method was not used for CCMC calculations in this study.
\begin{table*}
	\caption{Summary of complete basis set extrapolated results for the correlation
	energy of the 14 electron uniform electron gas in hartrees.}
	\begin{ruledtabular}
		\begin{tabular}{l|lllll}
			& $r_{\mathrm{s}}$ = 0.5 $\mathrm{a}_\mathrm{0}$ & $r_{\mathrm{s}}$ = 1.0
			$\mathrm{a}_\mathrm{0}$&$r_{\mathrm{s}}$ = 2.0
			$\mathrm{a}_\mathrm{0}$&$r_{\mathrm{s}}$ = 3.0
			$\mathrm{a}_\mathrm{0}$&$r_{\mathrm{s}}$ = 5.0 $\mathrm{a}_\mathrm{0}$\\
			\hline
			CCSD &-0.58850(6)/-0.5897(1)\footnote{This (initator) CCSD/CCSDT value is
			from Spencer et al.\cite{Spencer2016}} &
			-0.51450(9)/-0.5155(3)\footnotemark[1]\footnote{Also compare to -0.5152(5)
			from figure 7 in Shepherd et al.\cite{Shepherd2012} as quoted by Spencer et
			al. \cite{Spencer2016}}& -0.4096(10)/-0.4094(1)\footnotemark[1] & -0.3395(1)
			& -0.2531(3) \\
			CCSDT &-0.59457(7)/-0.5965(2)\footnotemark[1] &
			-0.5307(2)/-0.5317(3)\footnotemark[1] &
			-0.4407(10)/-0.4354(4)\footnotemark[1] & -0.3780(3)\footnote{The CCSDT to
			CCSD energy difference for $r_{\mathrm{s}}$ = 3.0 $\mathrm{a}_\mathrm{0}$ was
			estimated by the mean of a constant, linear, quadratic and cubic fit with
			lowest $\chi^2/$\#d.o.f. if multiple fits were available.} &
			-0.2970(4)\footnote{The CCSDT to CCSD energy difference for $r_{\mathrm{s}}$
			= 5.0 $\mathrm{a}_\mathrm{0}$ was estimated by the mean of a constant, linear
			and quadratic fit with lowest $\chi^2/$\#d.o.f. if multiple fits were
			available.} \\
			CCSDTQ & -0.59465(8) & -0.5311(2) & -0.4432(10)\footnote{The CCSDTQ to CCSDT
			difference for $r_{\mathrm{s}}$ = 2.0 and 3.0 $\mathrm{a}_\mathrm{0}$ and the
			CCSDTQ5 to CCSDTQ difference for $r_{\mathrm{s}}$ = 0.5
			$\mathrm{a}_\mathrm{0}$ was estimated by the mean of a linear fit and the
			data point with lowest $1/M$.} & -0.3833(3)\footnotemark[3]\footnotemark[5] &
			\textit{-0.3015(4)}\footnotemark[4]\footnote{The CCSDTQ to CCSDT difference
			for $r_{\mathrm{s}}$ = 5.0 $\mathrm{a}_\mathrm{0}$ and the CCSDTQ5 to CCSDTQ
			difference for $r_{\mathrm{s}}$ = 1.0, 2.0, 3.0 and 5.0
			$\mathrm{a}_\mathrm{0}$ was estimated by the CCSDTQ to CCSDT difference at 66
			spinorbitals.}\\
			CCSDTQ5& -0.5947(2)\footnotemark[5] & \textit{-0.5311(2)}\footnotemark[6] &
			\textit{-0.4434(10)}\footnotemark[5]\footnotemark[6] &
			\textit{-0.3837(3)}\footnotemark[3]\footnotemark[5]\footnotemark[6] &
			\textit{-0.3025(4)}\footnotemark[4]\footnotemark[6] \\ \hline
			FCIQMC & -0.59467(9)\footnote{(\textit{i})FCIQMC value of $r_{\mathrm{s}}$ =
			0.5 $\mathrm{a}_\mathrm{0}$ was estimated by the CCSDTQ value plus the
			difference of CCSDTQ to (\textit{i})FCIQMC extrapolated
			value.}/-0.5969(3)\footnote{This \textit{i}FCIQMC data is from Shepherd et
			al. \cite{Shepherd2012b}}& -0.5313(2)\footnote{(\textit{i})FCIQMC value of
			$r_{\mathrm{s}}$ = 1.0 $\mathrm{a}_\mathrm{0}$ was estimated by the CCSDT
			value plus the difference of CCSDT to (\textit{i})FCIQMC extrapolated
			value.}/-0.5325(4)\footnotemark[8]& -0.4447(4)\footnotemark[8]& &
			-0.306(1)\footnotemark[8]\\ \hline
			MP2& -0.575442(1)\footnote{The MP2 data is from Shepherd et al.
			\cite{Shepherd2012}}&
			-0.499338(2)\footnotemark[10]&-0.398948(2)\footnotemark[10]& &
			-0.255664(4)\footnotemark[10]
		\end{tabular}
	\end{ruledtabular}
	\label{tab:summary}
\end{table*} 

\section{Discussion}
Figure \ref{fig:energydiffs1} shows that CCSD gives an accuracy worse than 0.01
eV/electron for $r_{\mathrm{s}}$ greater than 0.5 $\mathrm{a}_\mathrm{0}$ as the
difference between CCSD and CCSDT is greater than 0.01 eV/electron. Considering
figure \ref{fig:energydiffs2}, CCSDT seems to be sufficient up to
$r_{\mathrm{s}}$ = 2.0 $\mathrm{a}_\mathrm{0}$. As the differences in
correlation energy increase in magnitude with $M$ and the $M$ = 162 energy for
$r_{\mathrm{s}}$ = 3.0 $\mathrm{a}_\mathrm{0}$ is close to 0.01 eV/electron, one
should be cautious about using CCSDT for $r_{\mathrm{s}}$ = 3.0
$\mathrm{a}_\mathrm{0}$. Figure \ref{fig:energydiffs3} shows that the difference
between \mbox{CCSDTQ} and \mbox{CCSDTQ5} is not negligible for $r_{\mathrm{s}}$
greater than 2.0 $\mathrm{a}_\mathrm{0}$. \par Of course, this analysis
implicitly assumes that the energy is monotonically decreasing with coupled
cluster level. If the difference to the next excitation level is bigger than
0.01 eV/electron, we expect the difference to the true energy also to be greater
than 0.01 eV/electron. However, we found that in our case, the energy was
monotonically decreasing and the CCSDTQ5 result agrees very well with FCIQMC,
see table \ref{tab:summary}. This supports our approach of comparing the energy
difference to the next excitation level when assessing accuracies.
\par
\begin{figure}
\centering
	\includegraphics[width=8.5cm,keepaspectratio]{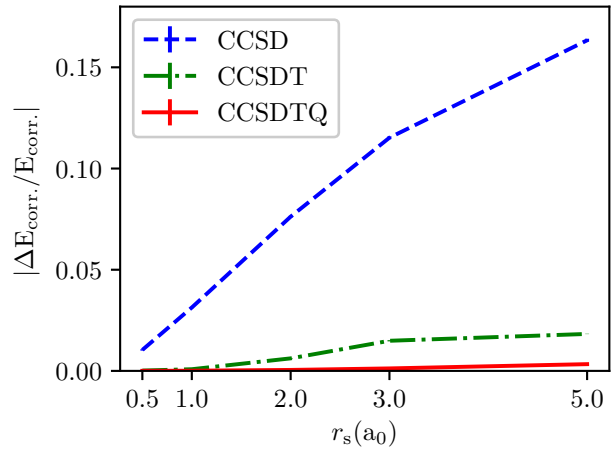}
	\caption{Fractional difference of CCSD, CCSDT and CCSDTQ correlation energies
	to the CCSDTQ5 correlation energy as a function of $r_{\mathrm{s}}$. Some
	coupled cluster correlation energies were estimated as described in table
	\ref{tab:summary}.}
	\label{fig:fraction}
\end{figure}

 Figure \ref{fig:fraction} shows the difference in correlation energy found
with CCSD, CCSDT and CCSDTQ to the correlation energy found with CCSDTQ5 as a
fraction of the CCSDTQ5 correlation energy. Given that the \mbox{CCSDTQ5} energy
shown in table \ref{tab:summary} is merely a lower bound for the true magnitude
of the CCSDTQ5 energy, the errors presented here are also lower bounds. The
error in CCSD is at least 16\% for $r_{\mathrm{s}}$ = 5.0
$\mathrm{a}_\mathrm{0}$ and for CCSDT it is still as big as about 2\%. The error
of CCSDTQ is small but noticable for $r_{\mathrm{s}}$ = 5.0
$\mathrm{a}_\mathrm{0}$. This means that for a study of a solid with
$r_{\mathrm{s}} \approx$ 4 $\mathrm{a}_\mathrm{0}$ say, e.g. sodium, CCSD may
give a correlation energy that is off by over 12\% and the error with CCSDT is
still over 1\%. As the energy differences between coupled cluster levels
increase with $r_{\mathrm{s}}$, properties such as the lattice parameter or the
bulk modulus will be underestimated by low orders of coupled cluster.
\par As Shepherd et al.\cite{Shepherd2012} already noted, for low
$r_{\mathrm{s}}$, MP2 performs worse than CCSD and vice versa for higher
$r_{\mathrm{s}}$ in the regime studied (see table \ref{tab:summary}). MP2 gives
a less accurate answer than CCSDT and higher truncation levels for all studied
$r_{\mathrm{s}}$.
\par We present new extrapolated FCIQMC results for $r_{\mathrm{s}}$ = 0.5 and
1.0 $\mathrm{a}_\mathrm{0}$, which are similar to but do not agree with Shepherd
et al.'s \cite{Shepherd2012b} values. Similarly, our CCSD and CCSDT values for
$r_{\mathrm{s}}$ = 0.5 and 1.0 $\mathrm{a}_\mathrm{0}$ do not agree within
2$\sigma$ with Spencer et al.'s \cite{Spencer2016} values. Our CBS correlation
energies are less negative. We can explain these deviations by considering the
shape of the extrapolation curves such as figure \ref{fig:extrapolation}. Our
CCSD calculations went up to 18342/11150 spinorbitals for $r_{\mathrm{s}}$ =
0.5/1.0 $\mathrm{a}_\mathrm{0}$ and that was our starting point to extrapolate
higher truncations and FCIQMC from. Shepherd et al. \cite{Shepherd2012b} and
Spencer et al. \cite{Spencer2016} only considered $M$ up to 4218 at most. If
fewer data points with low $1/M$ are present and a linear fit is employed (as
Shepherd et al. \cite{Shepherd2012b} and Spencer et al. \cite{Spencer2016} did),
the intercept with the $y$ axis, the CBS energy estimate, will be more negative
than in the case where lower $1/M$ are present and higher fits are allowed. Our
FCIQMC values quoted in table \ref{tab:summary} were found by extrapolating the
difference between the CCSDTQ/CCSDT and the FCIQMC values for $r_{\mathrm{s}}$ =
0.5/1.0 $\mathrm{a}_\mathrm{0}$ as CCSDTQ/CCSDT was the highest coupled cluster
data set that contained the highest $M$ used in our FCIQMC study for
$r_{\mathrm{s}}$ = 0.5/1.0 $\mathrm{a}_\mathrm{0}$ respectively. Had we instead
extrapolated FCIQMC directly, the results would have been -0.59497(4) ha
(instead of -0.59467(9) ha) with a linear fit for $r_{\mathrm{s}}$ = 0.5. For
this direct fit we included spinorbitals up to $M$ = 4218 and when we
extrapolated differences, we used information from the CCSD result with
spinorbitals up to 18342. This shows that it is crucial to include large numbers
of virtual orbitals to converge to the correct answer. We believe that the
disagreement of the CCSD and CCSDT values for $r_{\mathrm{s}}$ = 0.5 and 1.0
$\mathrm{a}_\mathrm{0}$ with Spencer et al.'s \cite{Spencer2016} values may also
be due to initiator energies that are not converged fully. We have not used the
initiator approximation for coupled cluster data here.

\section{Summary and Conclusions}
We have shown that CCSD and CCSDT are limited for modelling finite solids that can be
described by the 14 electron uniform electron gas with  $r_{\mathrm{s}}$ greater than 2.0
$\mathrm{a}_\mathrm{0}$. A comparison with CCSDTQ5 has shown that if an accuracy
of 0.01 eV/electron is desired, CCSDT is required beyond $r_{\mathrm{s}}$ = 0.5
$\mathrm{a}_\mathrm{0}$ and CCSDTQ is worth considering beyond $r_{\mathrm{s}}$
= 3.0 $\mathrm{a}_\mathrm{0}$. At $r_{\mathrm{s}}$ = 5.0
$\mathrm{a}_\mathrm{0}$, CCSD only reproduces up to about 84\% of the
correlation energy and CCSDT up to about 98\%.
\par This study has demonstrated that there can be a need for coupled cluster orders
beyond CCSDT when modelling finite correlated solid-state systems.

\begin{acknowledgments}
We thank Dr. Pablo L\'opez R\'ios for helpful discussions on fitting, especially
the idea of using $\chi^2/$(number of degrees of freedom).
For the research data supporting this publication and more information, see
https://doi.org/10.17863/CAM.14336.
V.A.N. would like to acknowledge the EPSRC Centre for Doctoral Training in
Computational Methods for Materials Science for funding under grant number EP/L015552/1.
A.J.W.T. thanks the Royal Society for a University Research Fellowship under
grant UF110161.
This work used the UK Research Data Facility
(http://www.archer.ac.uk/documentation/rdf-guide) and ARCHER UK National
Supercomputing Service (http://www.archer.ac.uk) under ARCHER Leadership project
with grant e507.
\end{acknowledgments}

\appendix


%

\end{document}